\documentclass{article}

\usepackage[T1]{fontenc}
\usepackage[utf8]{inputenc}
\usepackage{lmodern}
\usepackage[margin=1in]{geometry}

\usepackage{amsmath,amssymb,amsthm}
\usepackage{graphicx}
\usepackage{xcolor}
\usepackage{natbib}
\usepackage{hyperref}
\usepackage[nameinlink]{cleveref}

\hypersetup{
  pdftitle={Capture history analysis for spatial wildlife surveys with detection times},
  pdfauthor={Benjamin R. Baer, David L. Borchers, Greg Distiller},
  pdfkeywords={Spatial capture-recapture, survival analysis, removal surveys, single-catch traps},
  hidelinks
}

\newtheorem{prop}{Proposition}
\newtheorem{lem}{Lemma}
\newtheorem{cor}{Corollary}
\newtheorem{thm}{Theorem}
\newtheorem{asp}{Assumption}

\crefname{asp}{assumption}{assumptions}

\newcommand{\ind}{\perp\!\!\!\!\perp}

\renewcommand{\d}{\mathrm{d}}

\title{Capture history analysis for spatial wildlife surveys \\ with detection times}
\author{Benjamin R. Baer\\
  Department of Mathematics, Bowdoin College, Maine, USA\\
  and\\
  David L. Borchers\\
  Centre for Research into Ecological and Environmental Modelling,\\
  School of Mathematics and Statistics, University of St Andrews, Scotland\\
  and\\
  Greg Distiller\\
  Center for Statistics in Ecology, the Environment and Conservation,\\
  Department of Statistical Sciences, University of Cape Town, South Africa}
\date{}

\begin{document}
\maketitle

\begin{abstract}
    Many wildlife surveys of species with individually identifiable animals are based on the analysis of the capture histories of individual animals. However, not many of these take account of both the animals' capture times and their capture locations. 
    In this work, we develop a maximum likelihood estimator for surveys of individually identifiable animals using proximity detectors, multi-catch traps, single-catch traps, and traps that remove animals from the population, when capture times are known. We do this using the counting process theory at the foundation of event history (or survival) analysis. 
    The work provides a unifying framework for such surveys,  resolves open problems for single-catch traps and removal surveys, and introduces a new statistical method for multi-catch traps. 
    We test the new methods by simulation and we present an analysis of possums in New Zealand caught with single-catch traps. 
\end{abstract}

\noindent%

\section{Introduction}
\label{s:introduction}

Survival (or event history) models have found many uses in wildlife population assessment. They are typically used to model wildlife survival and mortality, but they have also been useful for modeling detection or capture processes on wildlife surveys. Examples of their use in this way include use in removal surveys \cite[e.g.,][]{zippin1958evaluation}, in distance sampling surveys \cite[][for example]{Schweder1974, HayesBuckland1983}, and more recently in models that combine distance sampling and capture-recapture models \citep[see][p. 595 for references]{BorchersCox2017}, as well as in spatial capture-recapture (SCR) models \citep[e.g.][]{BorchersEfford2008, EffordEtAl2013}.

Using survival functions to model the detection or capture process on a wildlife survey involves viewing detections or captures as ``deaths'', whether or not capture actually kills animals or removes them from the population by other means. (For brevity, we will from now on sometimes use the word ``capture'' to refer to both detections and physical ``captures'', and we will use the word ``trap'' to refer to both detectors and traps that hold animals.) Wildlife surveys consist of placing multiple traps in some survey region and recording which animals are detected by which traps and at what times. Not all surveys record detection times, but here we consider only those that do.

When traps hold or kill animals, one can model each individual’s time to capture using a competing-risks survival model, with traps ``competing''  to catch animals \citep[see][p. 379, for example]{BorchersEfford2008}. When captured animals are removed from the population, by being killed, or otherwise, the survey is a kind of removal survey \citep[see][for a description of the method]{zippin1958evaluation}. When animals are returned to the population after capture, and are also individually identifiable, one can view the survey as a spatial capture-recapture survey.

While maximum likelihood inference methods exist for both capture-recapture and removal methods, 
existing removal methods for estimating abundance do not account for the location of captures, and hence they do not account for spatial heterogeneity in capture probability. And maximum likelihood inference for abundance from SCR surveys using traps that are taken out of action by the act of capture (because the trap door shuts, for example) has proved intractable \citep[][p. 1530]{efford2019fast}. This is because of the difficulty of evaluating the survival function in the presence of stochastic trap removals, when each removal changes the total hazard of capture. 

Standard competing risks models in survival analysis assume the set of active risks is fixed, allowing the survival function to be obtained in closed form from the cumulative hazard \citep[e.g.][]{kalbfleisch2002}. Models with time-dependent covariates or shared frailties \citep[see, e.g.,][]{hougaard2000} introduce heterogeneity, but the hazard for any given individual depends only on that individual's own characteristics, so the survival function remains tractable. In the single-catch trap setting (when traps close upon catching an animal), by contrast, the hazard for each animal depends on the random capture outcomes of other animals and traps so that each closure removes a competing risk for all uncaptured animals, coupling the fates of individuals through the shared trap resource. This places the problem outside the scope of classical survival methods that treat individuals as fully independent and motivates the counting process approach we develop in \Cref{likelihood} \citep[cf.][]{andersen1993statistical}.  

Ideas from survival models have been used to tackle the single-catch trap problem in the past. \citet{distiller2015singlecatch} developed an approximate maximum likelihood estimator modeling capture as a survival process, for the case in which capture times are known. They fell short of developing a proper maximum likelihood estimator because of the difficulty of evaluating the survival function; their estimator relies on an approximation of this function. \citet{efford2004density} developed an inverse-prediction method for such single-catch trap SCR survey data, which does not require capture times and is not based on survival theory ideas. 

Here we adapt survival model theory to develop spatially explicit maximum likelihood methods for estimating abundance from spatial removal (SR) and spatial capture-recapture (SCR) surveys. In SR surveys, captured animals are removed from the population for the remainder of the survey. For SCR surveys, we consider single-catch traps, which can hold only one animal at a time; multi-catch traps, which can hold several animals at once; and proximity detectors, which record detections without holding animals. Animals caught in single- or multi-catch traps during an SCR survey are later returned to the population. 
The SCR methods require that animals are individually identifiable.

The remainder of this paper is organized as follows. In \Cref{likelihood} we use counting process theory to develop a general likelihood framework that covers all of the above cases and we study single-catch traps and removal surveys in detail. A simulation study is presented in \Cref{simulations}, in which we compare the performance of the maximum likelihood estimators (MLEs) for the SR method and SCR methods using single-catch traps, multi-catch traps, and proximity detectors. We also compare the performance of our MLE for a single-catch SCR survey with two existing methods that are commonly used to estimate density from this kind of survey, namely a multi-catch trap MLE and the inverse-prediction method of \citet{efford2004density}. In \Cref{application} we use our MLE for single-catch trap SCR data to analyze a survey of possums in New Zealand. We conclude with a Discussion in \Cref{discussion}.

\section{Methodology}
\label{sec:meth}

\subsection{Setup}
\label{sec:meth:setup}

Consider animals $i=1, \dots, N$ and traps $k=1, \dots, K$ in an environment $\mathcal{E}$. Define the counting process 
\begin{equation*}
    C_{ik}(t)
    := \# \{ \text{detections of animal } i \text{ at trap } k \text{ during } (0,t] \},
\end{equation*}
defined over $(0,\tau]$, where $\tau$ denotes the end of the study. 
Then, define the vector process $\pmb{C}(t) := \bigl( C_{11}(t), \dots, C_{NK}(t) \bigr)$ which gives the capture history of all animals, both observed and unobserved, and define the \emph{full data} as the path of $\pmb{C}$. 
Under the natural filtration $\mathcal{F}^*_t$ of the full data and in the absence of explosions, the Doob-Meyer theorem
gives a compensator 
\begin{equation*}
    \pmb{A}(t; \mathcal{F}^*_{t-}) = \bigl( A_{11}(t; \mathcal{F}^*_{t-}), \dots, A_{NK}(t; \mathcal{F}^*_{t-}) \bigr).  
\end{equation*}
Additionally, we specify that no two animals can be caught at the same time and no two traps can capture at the same time, which always holds when roughly speaking captures occur in ``continuous time''. 
\begin{asp}
\label{asp:nsj}
    The components of $\pmb{C}$ have no shared jumps.
\end{asp} 
\noindent 
Under \Cref{asp:nsj}, the vector process $\pmb{C}$ is a multivariate counting process, which allows the
standard tools of counting process theory to be employed. %

Define $C_{i\bullet}(t) := \sum_{k=1}^K C_{ik}(t)$ and $C_{\bullet k}(t) := \sum_{i=1}^N C_{ik}(t)$ as the total numbers of captures of animal $i$ and 
captures at trap $k$ in $(0,t]$, respectively. 
Define last-reset processes for animals $R_{A,i}(t)=\sup\{r<t:\text{ animal }i\text{ was reset at }r\}$ and 
for traps $R_{T,k}(t)=\sup\{r<t:\text{ trap }k\text{ was reset at }r\}$, with the convention that the supremum
is zero when the set is empty. For a trap, this is the most recent time at which it was reset and made available for capture; 
for an animal, it is the most recent time at which it was released from a trap and returned to the general population. For this purpose, we regard time $0$ as the initial reset time for every trap and animal.

\begin{table}[ht]
\caption{At-risk processes for common survey methods and detector types.}
\label{tab:atrisk}
\centering
\begin{tabular}{ll}
    \hline
    Survey method & At-risk process \\
    \hline
    Proximity & $1$ \\
    Multi-catch & $I\{C_{i\bullet}(t-) - C_{i\bullet}(R_{A,i}(t)) = 0\}$ \\
    Single-catch & $I\{C_{i\bullet}(t-) - C_{i\bullet}(R_{A,i}(t))=0\}$ %
    $\times I\{C_{\bullet k}(t-) - C_{\bullet k}(R_{T,k}(t))=0\}$ \\
    Removal & $I\{C_{i\bullet}(t-) = 0\}$ %
    $\times I\{C_{\bullet k}(t-) - C_{\bullet k}(R_{T,k}(t)) = 0\}$ \\
    \hline
\end{tabular}
\end{table}

Consider an at-risk process $Y_{ik}(t)=0,1$ that is $1$ when animal $i$ is available to be caught at 
trap $k$. 
Different survey methods and detector types will have different forms of $Y_{ik}$ in terms of the capture 
and reset histories. \Cref{tab:atrisk} shows the at-risk processes for each of the cases that we explicitly consider. 
For proximity detectors, each trap is always available to capture animals and each animal is always available to be caught. 
For single-catch traps, a trap can capture only one animal between resets, and an animal can be captured only once between releases. 
Removal surveys are similar to standard surveys with single-catch detectors but differ in that animals are never released. 
Let $E_{ik}(t)$ be a predictable eligibility process unrelated to animal or trap occupancy. 
For example, $E_{ik}(\cdot)$ can be used to track whether trap $k$ is undeployed. %
Formally, in these four cases, we define the at-risk process $Y_{ik}(t)$ as the expression in the table, which codifies 
the detector type, multiplied by the eligibility process $E_{ik}(t)$. Thus $Y_{ik}(\cdot)$ depends algebraically only on the capture histories and eligibility processes. 
Throughout, all probabilities, expectations, and likelihoods are conditional on the observed eligibility processes and we suppress this conditioning from the notation.

We now state the multiplicative intensity assumption, introduced by \citet{aalen1975phd} while providing
a comprehensive framework for survival analysis. 
As a general 
matter, the assumption does not specify the form of the at-risk process. 
\begin{asp}
\label{asp:mh}
    $A_{ik}(t; \mathcal{F}^*_{t-}) 
    = \int_{(0,t]} Y_{ik}(u) \, \d\Lambda_{ik}(u)$
    for each $k=1, \dots, K$ and $i=1, \dots, N$. 
\end{asp}
The assumption says that the capture intensity of animal $i$ and trap $k$ only depends on time
after taking into account the at-risk process. 
It fails, for example, when trap malfunctions are caused by unmeasured animal activity. Then eligibility is informative; for example, among traps that are currently active, traps with a history of ineligibility may have higher capture hazards than otherwise similar traps without such a history.

Another case in which \Cref{asp:mh} does not hold is when animals have post-capture behavioral changes. In such a case, two animals currently available for capture may have
different capture intensities when they have different capture histories. 
\citet{panchaud2024incorporating} consider a related relaxation of \Cref{asp:mh}, allowing the capture intensity to depend on both the time elapsed since the previous capture and the distance to the trap at which that capture occurred.
Although we make the multiplicative intensity assumption central to our development, in large part due to the simple and interpretable formulas it yields, analogous formulas can be derived under more general history-dependent intensities. Indeed, in \Cref{application}, we consider a modification of the intensity form that includes an additional term to model post-capture trap avoidance.

\subsection{Likelihood}
\label{likelihood}

In this section we derive the likelihood of the observed data. 

Let $n \leq N$ be the (random) number of captured animals. 
When animal $i$ is captured, the capture histories $C_{i1}(t), \dots, C_{iK}(t)$ across all traps are observed. 
For uncaptured animals, the capture histories are always zero since they were never detected. 
The observed data are thus the multiset 
\begin{equation*}
    \{ \pmb{C}_i(\cdot) \, : \, C_{i \bullet}(\tau) > 0, \, i = 1, \dots, N \},
\end{equation*}
which records the capture histories of captured animals and excludes the (unknown number of) zero capture histories.

Our first result gives the likelihood of the full data $\pmb{C}(\cdot)$. 
We use $\delta$ to denote the increment operator so that $\delta C_{ik}(t) := C_{ik}(t) - C_{ik}(t-)$ is the indicator for whether animal $i$ is captured at trap $k$ at time $t$. 

\begin{lem}
\label{lem:obs-lik}
    If \Cref{asp:nsj,asp:mh} hold, then the likelihood of the full data is 
    \begin{align*}
        & %
        \left\{ 
        \prod_{i=1}^n \prod_{k=1}^K \prod_{t\le \tau}
        \lambda_{ik}(t)^{\delta C_{ik}(t)}
        \right\} %
        \exp\left\{
        - \sum_{i=1}^N %
        \sum_{k=1}^K
        \int_0^\tau Y_{ik}(t) \d\Lambda_{ik}(t)
        \right\},%
    \end{align*}
    where $\lambda_{ik}(\cdot)$ denotes the density of $\Lambda_{ik}(\cdot)$. 
\end{lem}

\noindent The notation $\prod_{t\le \tau} \lambda_{ik}(t)^{\delta C_{ik}(t)}$ gives a product over the capture times of animal $i$ at trap $k$ of the capture hazard density $\lambda_{ik}(t)$.

In line with the development in \Cref{sec:meth:setup}, the lemma is general with respect to the at-risk processes and for example allows one animal's capture probability to depend on other animals' capture histories. Despite this dependence, there is \emph{sequential conditional independence} in light of \Cref{asp:mh}: for each instantaneous moment, an animal's capture probability at any trap is not influenced by the behavior of other animals or other traps in the same moment. The likelihood is then characterized incrementally by integrating each instantaneous moment; this is formally accomplished with Jacod's formula \citep{andersen1993statistical}. %

Consider activity centers $\pmb{s}_i$ which denote the ``typical'' locations of animals. The
distance $d_{ik} := \|\pmb{s}_i - \pmb{x}_k\|$ of animal $i$ to trap $k$, where $\pmb{x}_k$ is the position
of trap $k$, is known to be strongly related to the detection hazard. In the next assumption, we state that any heterogeneity in the
hazards across traps and animals is characterized by this distance. 
\begin{asp}
\label{asp:dist}
    $\Lambda_{ik}(t) = \Lambda(t; d_{ik})$. 
\end{asp}
\noindent %
Let $d_k(s)$ denote the distance of trap $k$ to activity center $s$. 

The next assumption states that the activity centers are independent and identically distributed, which makes
possible learning about the unobserved animal locations from the observed animal locations. 

\begin{asp}
\label{asp:iid}
    $\pmb{s}_1, \dots, \pmb{s}_N$ are i.i.d. and have Lebesgue density $f$. 
\end{asp}

\noindent In recent work, \citet{seaton2026spatialcapturerecapture} developed a method to relax this assumption for proximity detectors. 
We now treat the activity centers as included in $\mathcal{F}^*_0$. Thus \Cref{lem:obs-lik} gives the likelihood conditional on the activity centers. 

Additionally, we make the following technical assumption, which ensures that,
conditional on the full capture history, an animal's at-risk status can depend
on the activity centers only through its own activity center. It holds trivially in our setting, although it can be violated when each trap is accessible only to the animal whose activity center is closest to it, for example, because animals defend nonoverlapping territories. 

\begin{asp}
\label{asp:ac-sep}
    $Y_{ik}(t) \ind \bigl(\pmb{s}_1,\ldots,\pmb{s}_{i-1}, \pmb{s}_{i+1},\ldots,\pmb{s}_N\bigr) \mid \left(\pmb{s}_i, \{\pmb{C}(u):0\leq u<t\} \right)$ for all $i=1,\ldots,N$, $k=1,\ldots,K$, and $t\in(0,\tau]$. 
\end{asp}

Under these assumptions, we integrate out the unobserved activity centers to 
get a form of the likelihood that is recognizable relative to the standard
likelihoods that appear in the literature. 
Additionally, we integrate out all full data points compatible with the observed data
to derive the corresponding observed data likelihood; for a candidate $N \geq n$, every compatible full dataset assigns the $n$ observed nonzero histories to $n$ of the $N$ population labels and assigns zero histories to the remaining $N-n$ labels. 

\begin{thm}
\label{thm:lik}
    If \Cref{asp:nsj,asp:mh,asp:dist,asp:iid,asp:ac-sep} hold, then the likelihood of the observed data is 
    \begin{align*}
        & \binom{N}{n}
        \left[
        \int_{\mathcal{E}}
        \exp\left\{
        -\sum_{k=1}^K
            \int_0^\tau Y_{0k}(t)\d\Lambda\{t; d_{k}(\pmb{s})\}
        \right\}
        f(\pmb{s})\,\d \pmb{s}
        \right]^{N-n} \\
        & \hspace{10mm} \times \prod_{i=1}^n
        \Biggl[ \int_{\mathcal{E}}
        \left\{
            \prod_{k=1}^K \prod_{t\le \tau}
            \lambda\{t; d_{k}(\pmb{s})\}^{\delta C_{ik}(t)}
        \right\} %
        \exp\left\{
        -\sum_{k=1}^K
            \int_0^\tau Y_{ik}(t)\d\Lambda\{t; d_{k}(\pmb{s})\}
        \right\}
        f(\pmb{s})\,\d \pmb{s} \Biggr], 
    \end{align*}
    where $\lambda\{t; d_{k}(\pmb{s})\}$ denotes the density of $\Lambda\{t; d_{k}(\pmb{s})\}$
    and subscript $0$ denotes a generic unobserved animal with zero capture history. 
\end{thm}

The likelihood can be decomposed into binomial-looking
terms with a nice interpretation. Let $P_0$ denote the integral on the first line of the display in \Cref{thm:lik}, and let $P_i$ denote
the integral on the last line of the display in \Cref{thm:lik}. Then the likelihood in \Cref{thm:lik} admits the algebraic factorization 
\begin{align}
    & \binom{N}{n} \left( \prod_{i=1}^n P_i \right) P_0^{N-n} %
    =
    \left\{ \binom{N}{n} (1-P_0)^n P_0^{N-n} \right\} \prod_{i=1}^n \frac{P_i}{1-P_0}. \label{eq:binom}
\end{align}
The rightmost expression $\prod_{i=1}^n \frac{P_i}{1-P_0}$ resembles a conditional likelihood of the 
observed capture histories given that those animals were detected. The expression in curly braces 
$\binom{N}{n} (1-P_0)^n P_0^{N-n}$ resembles the binomial density of observing $n$ distinct animals in a population with abundance $N$. 
Despite the equation holding, we stress that we are not making binomial assumptions. For example, when there is only one single-catch trap and only one occasion, the binomial density in \Cref{eq:binom} is clearly misspecified since at most one animal (not all $N$ animals) can be captured. 

We implement maximum likelihood estimators by numerically maximizing the likelihoods in \Cref{thm:lik}. Following standard SCR practice, the integrals are numerically evaluated using a mesh \citep{BorchersEfford2008}. 
We report Wald standard errors and use a local normal approximation for interval estimation. 
Although the abundance $N$ is discrete, we treat it as continuous for estimation and inference. 

The likelihood in \Cref{thm:lik} is equivalent to the likelihood in \citet{borchers2014continuous}, in the case of proximity detectors. It appears to be novel in the case of the other surveys and detectors considered in \Cref{tab:atrisk}, and in the next two sections we consider single-catch and removal surveys in some detail.

\subsection{Single-catch traps}
\label{sec:meth:singlecatch}

\citet{distiller2015singlecatch} also model single-catch SCR data as a continuous-time competing-risks survival process and derive a likelihood, presented in their equation~(1). Because their likelihood includes a marginal detection probability that is computationally intractable, their proposed estimator instead maximizes a plug-in approximation. In this section, we describe their construction in our notation and show that the resulting plug-in approximation is a ``Poissonized'' version of the likelihood in our \Cref{thm:lik} and that, despite this support for their plug-in approximation, their likelihood derivation is based on an independence assumption that does not hold for single-catch traps. 

\begin{asp}
\label{asp:pois}
    $N \sim \mathrm{Poisson}(\mu)$. 
\end{asp}

Whenever we make this assumption, we interpret other assumptions that contain the abundance $N$ as being conditional on $N$. 

\begin{cor}
\label{cor:lik-pois}
    If \Cref{asp:nsj,asp:mh,asp:dist,asp:iid,asp:ac-sep,asp:pois} hold, then the likelihood of the observed data is 
    \begin{align*}
        & \frac{1}{n!}
        \exp\biggl[
        - 
        \int_{\mathcal{E}}\biggl\{
        1-
        \exp\left\{
        -\sum_{k=1}^K
            \int_0^\tau Y_{0k}(t)\,
            \d\Lambda\{t;d_k(\pmb{s})\}
        \right\}
        \biggr\}
        D(\pmb{s})\,\d\pmb{s}
        \biggr] \\
        &\hspace{5mm} \times
        \prod_{i=1}^n
        \biggl[
        \int_{\mathcal{E}}
        \left\{
            \prod_{k=1}^K \prod_{t\le \tau}
            \lambda\{t;d_k(\pmb{s})\}^{\delta C_{ik}(t)}
        \right\} %
        \exp\left\{
        -\sum_{k=1}^K
            \int_0^\tau Y_{ik}(t)\,
            \d\Lambda\{t;d_k(\pmb{s})\}
        \right\}
        D(\pmb{s})\,\d\pmb{s}
        \biggr].
    \end{align*}
    where $\lambda\{t; d_{k}(\pmb{s})\}$ denotes the density of $\Lambda\{t; d_{k}(\pmb{s})\}$
    and we denote $D(\cdot) := \mu f(\cdot)$. 
\end{cor}

The likelihood in \Cref{cor:lik-pois}, which is derived assuming that the abundance has a latent Poisson distribution, is exactly the plug-in approximation in \citet{distiller2015singlecatch} after correcting a typesetting omission where their survival function ``$S_{\bullet}$'' appears in place of ``$1 - S_{\bullet}$''. This places their likelihood-based inference on a conventional footing and helps explain its good empirical performance. 
Additionally, the leading terms in \Cref{cor:lik-pois} and \Cref{thm:lik} differ based on a law-of-rare-events approximation; in practice they are quite similar and their resulting maximizers are typically equal to practically important precision.

So far we have derived a likelihood under minimal assumptions and then showed that an additional Poisson assumption yields the plug-in approximation in \citet{distiller2015singlecatch}. This motivates a closer examination of the assumptions underlying the original likelihood formulation of \citet{distiller2015singlecatch}, which are not made fully explicit in that work.

Let $\pmb{C}_i = (C_{i1}, \dots, C_{iK})$ denote the capture history across traps for the $i$th animal and likewise let $\pmb{Y}_i = (Y_{i1}, \dots, Y_{iK})$ denote the at-risk processes across traps for the $i$th animal. 
\begin{asp}
\label{asp:ind-hist}
    (i) The processes $(\pmb{C}_i, \pmb{Y}_i)$ are conditionally independent across $i=1,\dots,N$ given $N$ and $\pmb{s}_1,\dots,\pmb{s}_N$, 
    (ii) their distributions depend on the activity centers only through $\pmb{s}_i$, and 
    (iii) $\pmb{Y}_i \mid \pmb{C}_i, \pmb{s}_i$ does not depend on $\pmb{s}_i$, the model parameters, or the animal $i=1, \dots, N$.
\end{asp}

The core component of \Cref{asp:ind-hist} is part (i), which specifies that the animals are independently captured and at risk of capture. 
Although \citet{distiller2015singlecatch} allow capture hazards to depend on the realized shared trap availability, their likelihood retains the independence structure used by \citet{borchers2014continuous} for proximity detectors. The next result derives their likelihood under this assumption. 

\begin{prop}
\label{prop:lik-pois2}
    If \Cref{asp:nsj,asp:mh,asp:dist,asp:iid,asp:ac-sep,asp:pois,asp:ind-hist} hold, 
    then the likelihood of the captured animal data is 
    \begin{align*}
        & \hspace{-10mm} \frac{1}{n!}
        \exp\biggl[
        - 
        \int_{\mathcal{E}}\biggl\{
        1-
        \mathbb{E} 
        \left[
        \exp\left\{
        -\sum_{k=1}^K
            \int_0^\tau Y_{0k}(t)\,
            \d\Lambda\{t;d_k(\pmb{s})\}
        \right\}
        \right]
        \biggr\}
        D(\pmb{s})\,\d\pmb{s}
        \biggr] \\
        &\hspace{3mm} \times
        \prod_{i=1}^n
        \biggl[
        \int_{\mathcal{E}}
        \left\{
            \prod_{k=1}^K \prod_{t\le \tau}
            \lambda\{t;d_k(\pmb{s})\}^{\delta C_{ik}(t)}
        \right\} %
        \exp\left\{
        -\sum_{k=1}^K
            \int_0^\tau Y_{ik}(t)\,
            \d\Lambda\{t;d_k(\pmb{s})\}
        \right\}
        D(\pmb{s})\,\d\pmb{s}
        \biggr], 
    \end{align*}
    where
    $\lambda_{ik}$ denotes the density of $\Lambda_{ik}$,
    we denote $D(\cdot) := \mu f(\cdot)$. 
\end{prop}

Unlike \Cref{cor:lik-pois}, which evaluates the undetected-animal term using the observed at-risk process, the likelihood in \Cref{prop:lik-pois2} averages this term over possible at-risk processes, including trap-closure histories other than the one observed. Roughly speaking, this averaging occurs because the independence in \Cref{asp:ind-hist} makes the at-risk process for uncaptured animals ``separate'' from that of the captured animals.

In a single-catch setting, the data $C_{ik}$ and $Y_{ik}$ do not generally satisfy \Cref{asp:ind-hist} since animal captures are intertwined through trap competition; see, e.g., \Cref{tab:atrisk}. 
For proximity and multi-catch surveys, \Cref{asp:ind-hist} does hold since the at-risk processes for undetected animals are deterministically equal to one during periods of active sampling; for these traps, although the expectation is present in the display, it is degenerate and does not create the difficulty that arises for single-catch data. 
However, due to the inappropriateness of the assumption for single-catch data, we do not need to entertain these difficulties.

\subsection{Removal surveys}

In this section, we consider the general likelihood in the case of a removal survey in which traps are single-catch and animals are removed from the population after their capture. 
The general likelihood in \Cref{thm:lik} has the binomial representation,
\begin{equation*}
     \left\{ \binom{N}{n} (1-P_0)^n P_0^{N-n} \right\} \prod_{i=1}^n \frac{P_i}{1-P_0}
\end{equation*}
also shown in \Cref{eq:binom}, where $P_i$ is the capture density for the $i$th captured animal and $P_0$ is the no-capture probability for an uncaptured animal. 

\citet{zippin1958evaluation} developed a maximum likelihood estimator for abundance using the same likelihood, but with different formulas for $P_i$ and $P_0$. Suppose momentarily that the traps are reset at discrete occasions, so that all traps are reset at times $\tau_1, \dots, \tau_m = \tau$. 
The likelihood from Zippin uses the probabilities
\begin{align*}
    P_i & = \exp (-\lambda \tau_{J_i-1}) \Bigl[ 1 - \exp \Bigl\{ - \lambda (\tau_{J_i} - \tau_{J_i-1}) \Bigr\} \Bigr] \\ 
    P_0 & = \exp ( - \lambda \tau_m ), 
\end{align*}
where $\lambda$ is the capture hazard and $J_i$ denotes the occasion when animal $i$ was captured. The expression for $P_i$ contains the term $\exp (-\lambda \tau_{j-1})$ giving the probability an animal was not caught up to time $\tau_{j-1}$ and the second term giving the probability an animal is caught between times $\tau_{j-1}$ and $\tau_j$, i.e., during occasion $j$. 
Only the capture occasion, rather than the exact capture time, must be known to calculate the likelihood, and the likelihood is critically derived under the assumption that all animals have the same capture hazard $\lambda$ during all occasions when they're at risk. 

This assumption may be unrealistic in ecological surveys because a trap becomes unavailable after capturing an animal, thereby reducing the capture intensity of the remaining animals. For example, the capture intensity is zero when all traps are closed. 
\citet{lin1999parametric,yip2007population} adapt the classical removal method to depend on the exact capture time (as we do) and allow for capture intensity to drop after each animal's capture. However, they assume that while a trap is available, every uncaptured animal has the same per-trap capture hazard at every trap; 
in our notation, they assume that the capture hazard $\lambda_{ik}(t) = \lambda$ doesn't depend on animal $i$, trap $k$, or time $t$. 

Although the assumptions of the method in \citet{yip2007population} are closer to our setting of spatial traps, this constant capture hazard assumption is still unrealistic in our setting: animals whose activity centers lie far from the traps generally have lower capture hazards. 
Using the same ideas as just presented for the \citet{zippin1958evaluation} likelihood, an improved likelihood uses the probabilities 
\begin{align*}
    P_i & = \exp\left\{
        - \sum_{k=1}^K
        \int_{(0,T_i)} Y_{ik}(t) \d\Lambda_{ik}(t)
        \right\} %
        \lambda_{ik_i}(T_i) \\
    P_0 & = \exp\left\{
        - \sum_{k=1}^K
        \int_0^\tau Y_{0k}(t) \d\Lambda_{0k}(t)
        \right\}, 
\end{align*}
where $T_i$ denotes the (single) capture time of the $i$th captured animal and $k_i$ denotes the trap that captured it. 
The terms in the density $P_i$ that animal $i$ was captured have the same interpretation as before; for example, the first term $\exp\left\{ - \sum_{k=1}^K \int_{(0,T_i)} Y_{ik}(t) \d\Lambda_{ik}(t) \right\}$ is the probability that animal $i$ was not captured before time $T_i$. However, now, this expression acknowledges the intensity changes depending on which traps are open or closed and additionally allows for the capture hazard to vary with trap, animal, and time.  

Indeed, at this point, the likelihood is the same as what is presented in \Cref{lem:obs-lik}. The development in \Cref{likelihood} then culminates in the presented likelihood, which hinges on \Cref{asp:dist} that $\Lambda_{ik}(t) = \Lambda(t; d_{ik})$ so that all trap-animal capture heterogeneity is determined by activity-center distance, and then integrates out activity centers. 
Under this assumption there is spatial information in the data, in that an animal being caught at a trap encourages its activity center to be close to that trap and away from the other traps.

Now, we show that a spatial removal survey identifies abundance through the instantaneous change in capture intensity after a removal. This follows the logic of classical removal methods such as \citet{zippin1958evaluation}, which infer abundance from changes in capture intensity as the population is depleted.
After integrating out the activity centers, all animals not yet captured have the same hazard at a given trap because they share the same zero capture history. The trap-level intensity is therefore the number of animals remaining multiplied by this common per-animal hazard. At the first capture, the former decreases from $N$ to $N-1$, while, at an unaffected trap, %
the latter has no instantaneous change. The resulting pre–post intensity ratio therefore isolates $N$, as formalized below. 

\begin{prop}
\label{prop:removal-identification}
    Consider a removal survey in which \Cref{asp:nsj,asp:mh,asp:dist,asp:iid,asp:ac-sep} hold, the capture hazard is right-continuous, and $N \geq 2$. If, after the first capture, at least one other trap remains at risk with positive conditional capture intensity, then the distribution of the observed data identifies the abundance $N$. 
\end{prop}

In addition to the abundance $N$, removal data may contain information about spatial detection parameters such as the scale $\sigma$ under a half-normal capture-hazard model. Rather than entering through the instantaneous intensity contrast as in the preceding proposition, these parameters determine how the trap-specific conditional intensities evolve over the survey. 
Animals with larger capture hazards tend to be caught earlier, and, because capture hazard depends on activity-center distance, the activity-center distribution of the uncaptured animals becomes increasingly weighted toward locations giving small cumulative hazards at the traps that have remained open. 
If capture hazard decreases rapidly with distance, this selective depletion is relatively local; if it decreases more slowly, its effects extend more broadly across the trap array. 
The subsequent capture times and locations, together with the observed pattern of trap closures and resets, may therefore be informative about the spatial reach of selective depletion.

Put differently, survey methods with recaptures reveal the spatial reach of detection directly when the same animal is captured at different traps; removal surveys instead contain information indirectly through changes in the capture process among the animals that remain.
In future work, we plan to investigate these identification questions further, with particular attention to distinguishing spatial detection effects from variation in the density surface and temporal variation in capture hazard.

\section{Simulation Study}
\label{simulations}

In this section, we study the maximum likelihood estimators based on the presented likelihoods in \Cref{thm:lik} in a simulation scenario based on \citet{distiller2015singlecatch}. In the setting of single-catch data, please see \citet{distiller2015singlecatch,efford2023ipsecr} for extended simulations and discussions. 
We consider an environment containing $20$ traps in a $5 \times 4$ grid with unit spacing 
and a buffer of length $4$ around the traps. 
Animal activity centers are independently sampled from the spatial surface $f_N(x):=\exp(\beta_0+\beta_1x)$, where $x$ is the horizontal position scaled to lie between $0$ on the left and $1$ on the right. 
We set $\beta_1=1.5\log(3)$ and vary $\beta_0$ to obtain mean densities of approximately $0.980$, $1.470$, $1.960$, $2.940$, and $5.881$, corresponding to abundances of $134$, $201$, $268$, $403$, and $806$, respectively.
We first set $N=\left\lfloor\int_{\mathcal{E}}f_N(\pmb{s})\,\mathrm{d}\pmb{s}\right\rfloor$. After rounding, we adjust $\beta_0$ so that $\int_{\mathcal{E}}f_N(\pmb{s})\,\mathrm{d}\pmb{s}=N$, and define the normalized activity-center density as $f=f_N/N$.
The true detection hazard function is $d\mapsto\lambda_0e^{-d^2/(2\sigma^2)}$, the usual half-normal form, where $\lambda_0=-\log(1-0.2)$ and $\sigma=1$. 
The study starts at time $0$ and ends at time $\tau=10$. %
All traps are reset at times $1,\ldots,9$, giving ten occasions of unit duration.

\begin{figure}[t]
    \includegraphics[scale=.65]{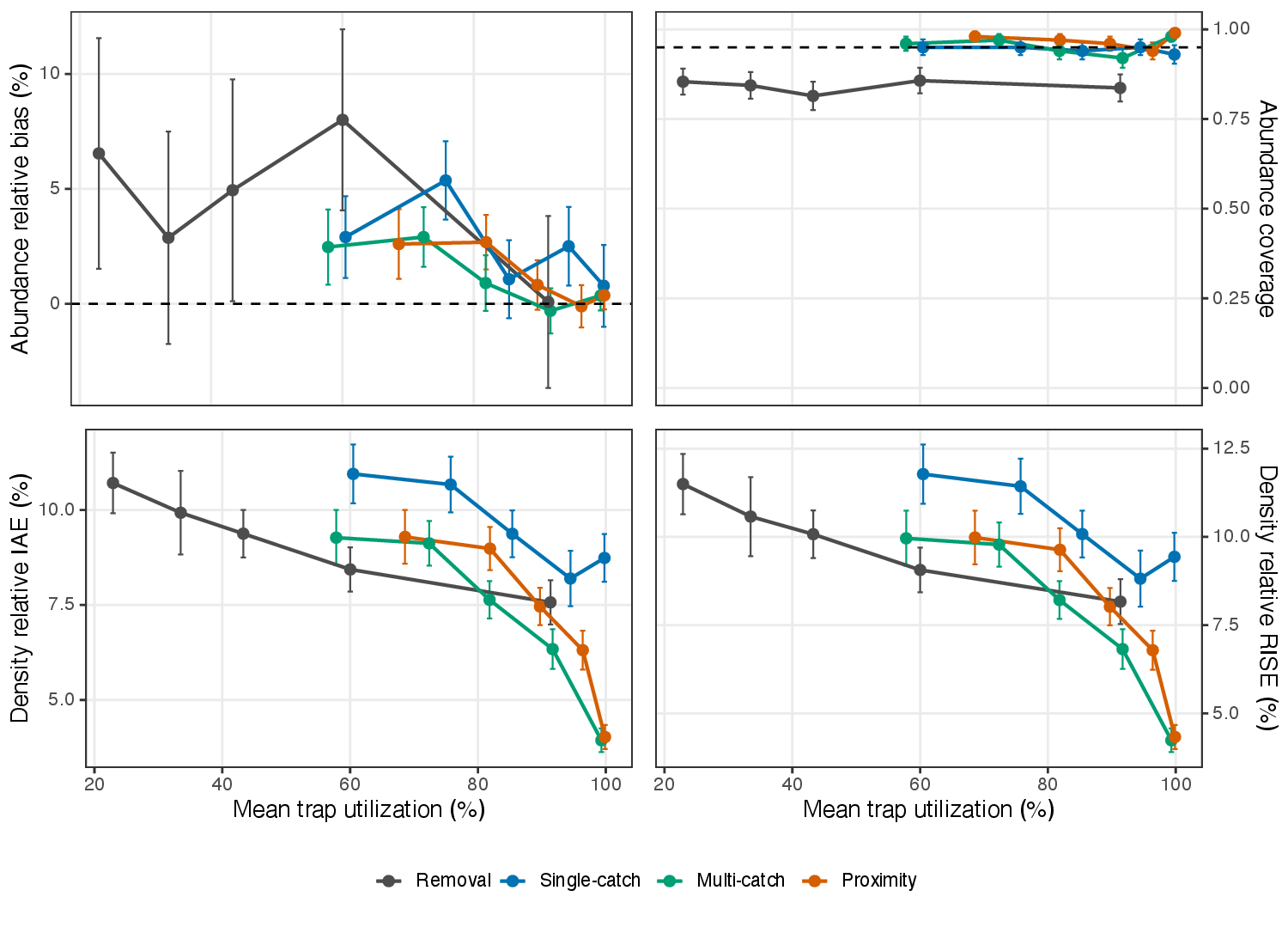}
    \caption{A comparison of the removal, single-catch, multi-catch, and proximity maximum likelihood estimators when each survey type is fitted using its correctly specified likelihood.
    }
\label{fig:coverage}
\end{figure}

\begin{figure}[t]
    \includegraphics[scale=.65]{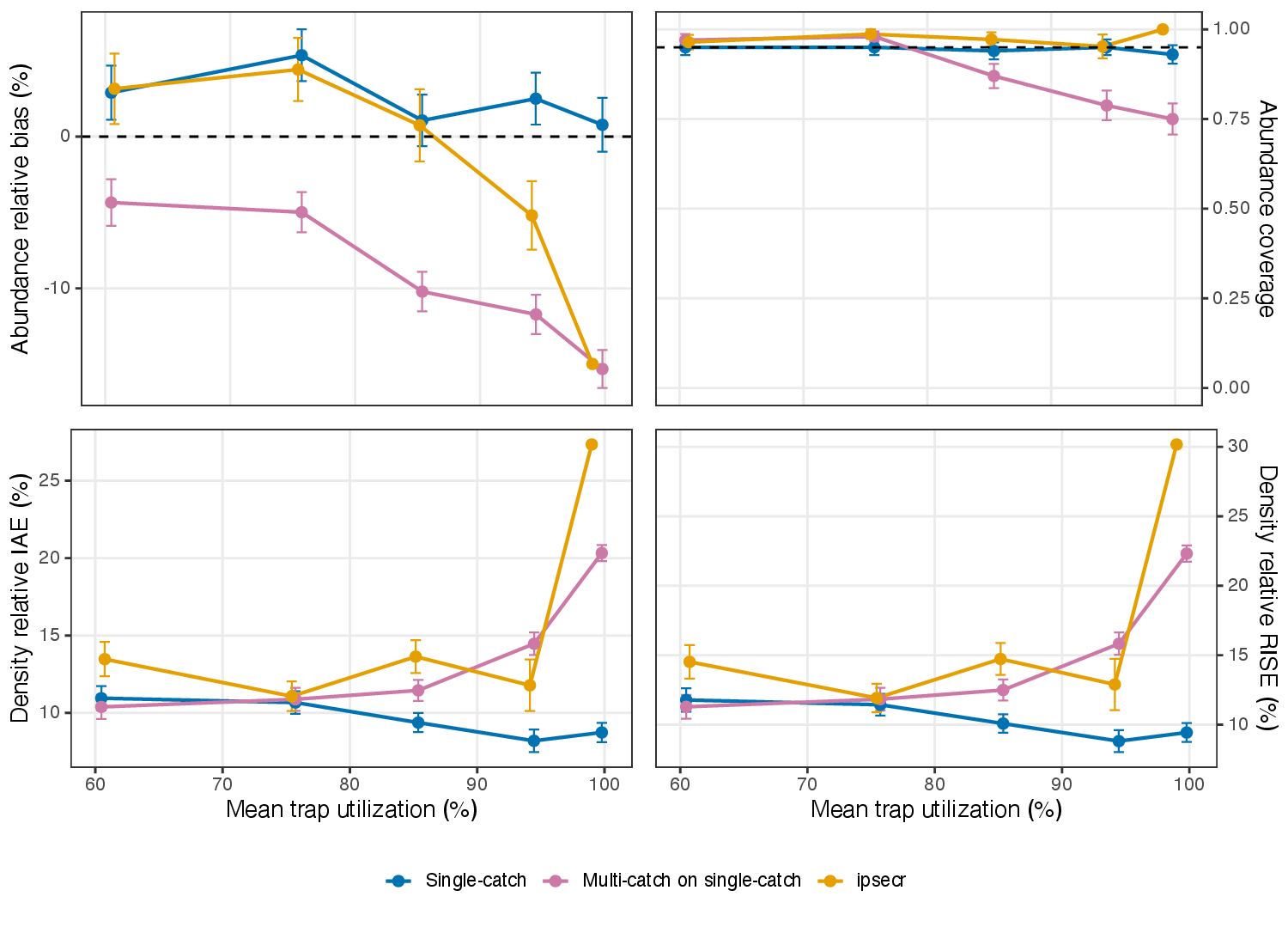}
    \caption{A comparison of the single-catch MLE, the misspecified multi-catch MLE, and the inverse-prediction estimator (among the $54.6\%$ of successful fits, decreasing to $1\%$ for high saturation) on single-catch data.}
\label{fig:singlecatch}
\end{figure}

\begin{figure}[t]
    \includegraphics[scale=.65]{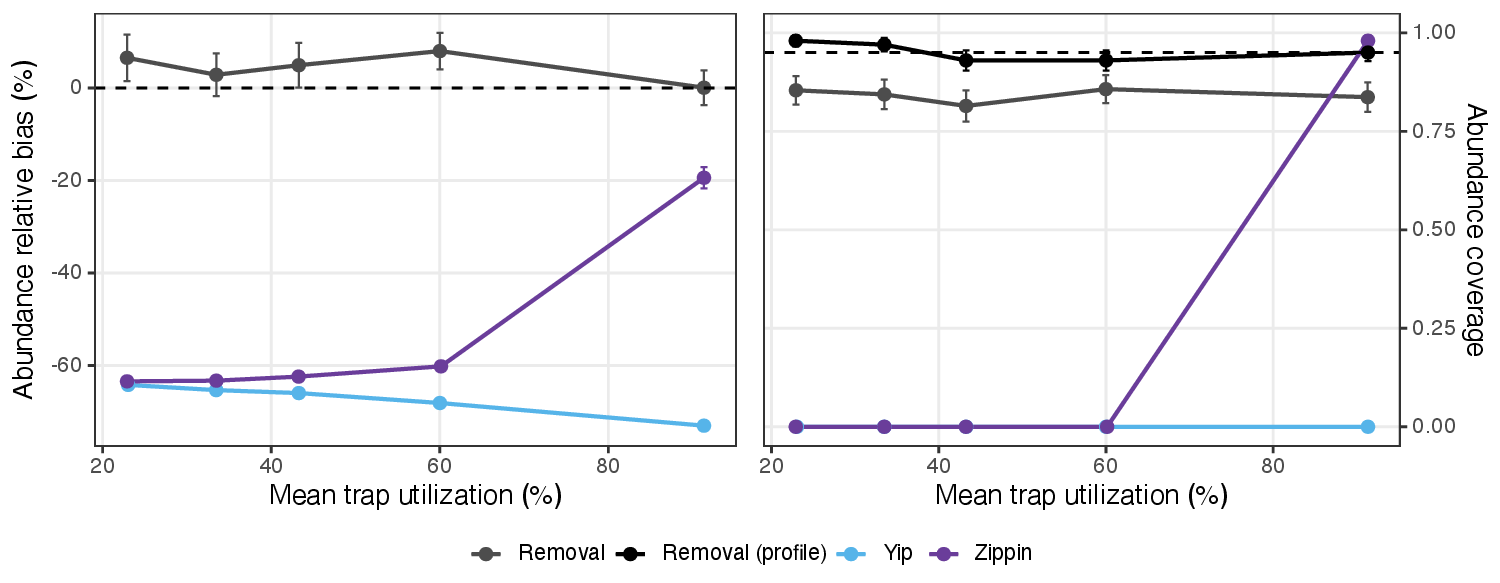}
    \caption{A comparison of the removal MLE and the non-spatial Zippin and Yip estimators on removal data.}
\label{fig:removal}
\end{figure}

For each animal-trap pair, potential capture times are generated by an independent Poisson process whose rate depends on activity-center distance. The observed data are then generated by retaining only potential captures for which the corresponding at-risk process in \Cref{tab:atrisk} equals one. Thus, no thinning occurs for proximity detectors, whereas in a removal survey a potential capture is discarded if the trap is already occupied or the animal has already been captured. 
In the simulation, varying the mean density produces a range of trap-utilization percentages, allowing us to examine estimator performance along a sequence of increasing mean density and resulting trap utilization, which we measure as the percentage of trap--occasion combinations in which at least one detection is recorded. 
We report the following four operating characteristics: the abundance relative bias, $100(\widehat{N}-N)/N$; the $95\%$ log-scale abundance coverage; the relative integrated absolute error (IAE), $100 \int_{\mathcal{R}}|\widehat{f}(s)-f(s)|\,\d s / \int_R f(s)\d s$; and the mean relative root integrated squared error (RISE) $100 [ \int_{\mathcal{R}}\{\widehat{f}(s)-f(s)\}^2 \d s/ \int_{\mathcal{R}} f(s)^2 \d s]^{1/2}$, 
where $\widehat{N}$ is an abundance estimate, $\widehat{f}(\cdot)$ is a density estimate, and $\mathcal{R}$ is the proper subset of the full environment $\mathcal{E}$ that only extends $2$ units beyond the trap convex hull. 
The plots in \Cref{fig:coverage,fig:singlecatch,fig:removal} show Monte Carlo averages of the operating characteristics of our estimators based on $100$ simulation replicates with one-standard-error bars.

\Cref{fig:coverage} shows the performance of the maximum likelihood estimators for the four survey types as mean trap utilization increases.
The top panels show the performance of the abundance estimators, while the bottom panels show the performance of the density-surface estimators. The upper-left panel shows abundance relative bias, the upper-right panel shows empirical coverage of the Wald $95\%$ confidence interval for $\log N$, and the lower panels show the relative density-surface integrated absolute error and relative root integrated squared error, respectively.
The abundance estimators for the proximity, multi-catch, and single-catch surveys have relatively little bias across the range of trap utilization. The removal estimator is more variable.
The empirical coverage of the Wald confidence interval for $\log N$ is generally close to the nominal level for the proximity, multi-catch, and single-catch estimators, ranging from approximately $92\%$ to $99\%$. Coverage is poorer for the removal estimator, ranging from approximately $81\%$ to $86\%$. The poor Wald coverage for abundance under the removal survey method may reflect weak identifiability.
For estimation of the density surface, proximity and multi-catch surveys perform similarly. The single-catch estimator has somewhat larger errors, particularly near complete trap utilization. The removal estimator generally has larger density-surface errors than the proximity and multi-catch estimators, especially at the higher-density settings, but performs similarly to the single-catch estimator across much of the simulation.

\Cref{fig:singlecatch} compares the proposed single-catch maximum likelihood estimator with two alternatives for the single-catch data. We do not include the maximizer of the approximate likelihood of \citet{distiller2015singlecatch}, studied in \Cref{sec:meth:singlecatch}, because it is theoretically and numerically similar to the proposed estimator.
The proposed estimator is shown in blue, the maximizer of the incorrectly specified multi-catch likelihood is shown in pink, and the inverse-prediction estimator of \citet{efford2023ipsecr}, which only uses the capture occasion rather than the exact capture time, is shown in orange.
The inverse-prediction estimator only produced an estimate on 84, 75, 71, 42, and 1 of the 100 replicates as trap utilization increased. The accompanying package documentation notes that estimation may fail and recommends case-specific manual tuning \citep{efford2023ipsecr}. Because such intervention was impractical in our setting across simulated replicates, we retained the default settings. Consequently, its operating characteristics in the figure should be interpreted cautiously.
The proposed single-catch estimator has relatively little abundance bias across the range of trap utilization, whereas the multi-catch estimator becomes increasingly negatively biased as trap utilization increases. Among successful fits, the inverse-prediction estimator also develops appreciable negative bias at high trap utilization. Wald coverage for the proposed estimator remains between approximately $93\%$ and $95\%$, while coverage for the multi-catch estimator declines to $75\%$.
The density-surface errors of the three estimators are most similar at lower trap utilization, while the proposed estimator performs appreciably better than both alternatives at higher utilization.

\Cref{fig:removal} compares the proposed spatial removal estimator, shown in gray, with the non-spatial Zippin estimator of \citet{zippin1958evaluation}, shown in purple, and the non-spatial Yip estimator of \citet{yip2007population}, shown in blue, on data from a removal survey. Because the non-spatial estimators do not estimate a density surface, only the abundance bias and coverage are shown.
The spatial estimator has substantially less abundance bias than the Zippin and Yip estimators. The non-spatial abundance estimators have bias approximately $-60\%$ over the first four density settings, with the bias of the Zippin estimator improving at the highest density setting. 
The Wald interval of the Zippin estimator and the Yip estimator have no empirical coverage over the first four settings, although coverage increases to, perhaps coincidentally, $98\%$ at the highest density for Zippin. Wald coverage for the spatial estimator is also below the nominal level, but its profile-likelihood interval, shown in black, attains approximately $93\%$--$98\%$ coverage. 
The apparent improvement of the Zippin estimator at the highest density may reflect cancellation between two sources of misspecification. 
Ignoring spatial heterogeneity tends to bias abundance downward, because low-detectability animals are mistaken for evidence that the population has been exhausted, whereas 
ignoring finite trap availability tends to bias abundance upward, because trap saturation keeps successive occasion totals artificially similar, which is interpreted as evidence that removals constitute only a small fraction of a large population. 
The Yip estimator accounts for trap availability and therefore removes this compensating upward bias, but remains susceptible to the downward bias induced by spatial heterogeneity.

\section{Application: Brushtail possums in New Zealand}
\label{application}

Common brushtail possums (\emph{Trichosurus vulpecula}) were introduced
to New Zealand in the nineteenth century to establish a fur industry
\citep{efford2004long-term}. Their selective browsing and predation on
indigenous birds and invertebrates have made them major pests
\citep{campbell1990changes,sadleir2000evidence}.

The possum study site was located in mixed podocarp-hardwood forest in
the Orongorongo Valley on the North Island of New Zealand; further site
details are given by \citet{efford2004long-term} and
\citet{cowan2012behavioural}. Possums were live-trapped on a grid of
wire-mesh cage traps spaced 30 m apart. Traps were baited with pieces of
apple coated in flour and mixed with aniseed oil. Trapping was conducted
for three consecutive days each month, and traps were checked and reset
daily. 

\begin{figure}
    \centering
    \includegraphics{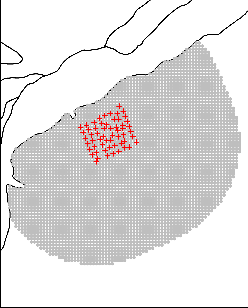} 
    \includegraphics[scale=.7]{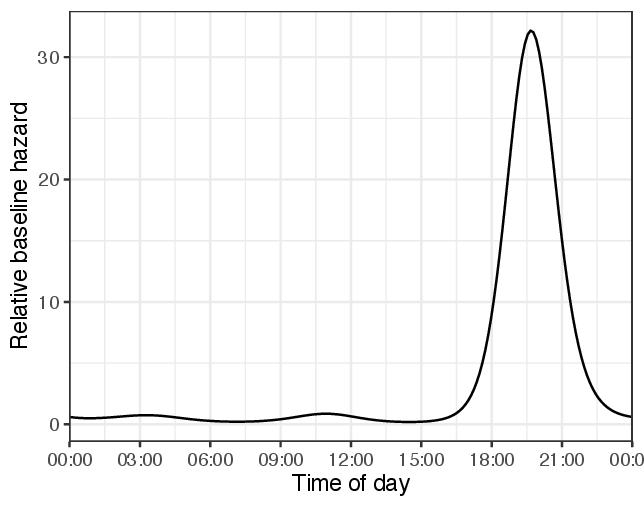}
    \caption{Left: the 79.9 ha survey area of the possum analysis, with traps denoted with red crosses. Right: the estimated relative capture hazard at distance zero.}
    \label{fig:possum}
\end{figure}

The trapping grid was bounded to the north and west by the Orongorongo
River. Apart from the river, habitat surrounding the grid was similar for
several kilometers. \Cref{fig:possum} shows the study area used in the possum analysis, with the river and the area across it excluded as potential locations for activity centers.

Timing devices, accurate to within five minutes, were attached to trap
door frames and activated when trap doors closed.
The five-minute timing precision resulted in some tied recorded times. No ties involved the same animal or trap, and we regard the resulting timing error as negligible.
Non-target species,
including rats (\emph{Rattus rattus}), mice (\emph{Mus musculus}), and
European hedgehogs (\emph{Erinaceus europaeus}), occasionally triggered
traps; in the analysis, the corresponding eligibility process is zero while a trap is unavailable after such a trigger. 

To maintain a closed-population assumption, the analysis used data from
August, September, and October 1982. These months avoid the main period
of juvenile independence and dispersal, which occurs mostly from October
to April, although some movement may also occur around the April-June
mating period. 

The data were received as handwritten timing sheets and
then transcribed and formatted for analysis. In the analyzed dataset,
70 unique possums were captured at 57 cage traps, producing 286 capture
events. The average trap saturation, defined as the mean proportion of
traps occupied at the end of an occasion, was 58\%.

The data were previously studied in \citet{distiller2016thesis}, an unpublished PhD thesis, where model selection was performed 
using the plug-in approximation reviewed in \Cref{sec:meth:singlecatch}. 
In this work, we use a similar model for detection in the presented likelihood in \Cref{thm:lik}, where the spatial density $D(\pmb{s}) = \exp (d_0)$ is flat and the cumulative hazard of capture is 
\begin{align*}
    & \Lambda\{t; d_k(\pmb{s}), C_{i\bullet}(t-)\} 
    = \int_0^t 
    \lambda_0 
    \exp\left\{ - \frac{d_k(\pmb{s})^2}{2\sigma^2} \right\} %
    \exp \biggl[ 
      \beta_0 I\{ C_{i\bullet}(u-) > 0 \} 
      + \beta_1\cos\left(\frac{2\pi u}{24}\right) \\
      & \hspace{10mm} 
      + \beta_2\sin\left(\frac{2\pi u}{24}\right) 
      + \beta_3\cos\left(\frac{4\pi u}{24}\right) 
      + \beta_4\sin\left(\frac{4\pi u}{24}\right) 
      + \beta_5\cos\left(\frac{6\pi u}{24}\right) %
      + \beta_6\sin\left(\frac{6\pi u}{24}\right) 
    \biggr] 
    \,\mathrm{d}u, 
\end{align*}
where time is measured in hours. 
The cumulative capture hazard combines a baseline detection rate, spatial decay with distance from the trap, a post-capture behavioral effect, and a daily activity pattern. Temporal variation is modeled with sine and cosine terms having a 24-hour period, allowing the capture hazard to vary over the course of each day.
Additionally, the term $\exp [ \beta_0 I\{ C_{i\bullet}(u-) > 0 \} ]$ makes the baseline hazard change from $\lambda_0$ to $\lambda_0 e^{\beta_0}$ after an animal is first captured. This is a violation of \Cref{asp:mh}, but, as discussed, the likelihood is still valid with this display included in the intensity since $C_{i\bullet}(t-)$ is predictable and Jacod's formula remains applicable. 

\begin{table}[ht]
\caption{Selected parameter estimates from the possum data analysis with standard errors in parentheses.}
\centering
\begin{tabular}{lcccc}
    \hline
    Parameter & $\lambda_0$ & $\sigma$ & $\beta_0$ & $d_0$ \\
    \hline
    Estimate & 0.006 (0.0011) & 37.88 (1.77) & -0.534 (0.171) & 1.73 (0.118) \\
    \hline
\end{tabular}
\label{tab:parameter-estimates}
\end{table}

Selected parameter estimates are shown in \Cref{tab:parameter-estimates} and the estimated temporal variation is depicted at the right of \Cref{fig:possum}. 
Combining the estimated density of $\exp (1.73) \approx 5.64$ possums per hectare with the survey area gives an estimated abundance of $\widehat{N}=453$ possums in the survey region. The estimated spatial scale, $\widehat{\sigma}=37.88$ m, is comparable with the 30 m trap spacing. The negative post-capture coefficient, $\widehat{\beta}_0 = -0.534$, corresponds to a multiplicative change of $\exp(-0.534)=0.59$ in baseline hazard after first capture, in line with trap avoidance. The fitted temporal effect is strongly nocturnal, with peak fitted relative activity occurring at approximately 19:30.

\section{Discussion}
\label{discussion}

The work of \citet{efford2009density} was the first in spatial capture-recapture to consider different detector types. They considered proximity detectors, multi-catch traps, and single-catch traps and presented statistical methods for proximity detectors and multi-catch traps, all in the case where only the capture occasion rather than the capture time is observed. 
In this work, we similarly consider different survey methods and detector types; however, via the counting process theory at the heart of event history analysis, we present a unified likelihood that encompasses their detectors and also many other possible survey methods and detector types.

Concerning single-catch traps, \citet{efford2009density} wrote ``a likelihood model for single-catch traps is considerably more complicated than for multi-catch traps, and remains to be developed.'' As previously reviewed, \citet{distiller2015singlecatch} presented a working likelihood for single-catch data and showed it to be analytically intractable due to the appearance of a high-dimensional expectation, after which the prevailing view in the statistical ecology community was that SCR with single-catch data was computationally intractable. 
For example, \citet{efford2023ipsecr} reviews and advocates for a computational workaround that we compared against in simulations. 
In this work, we showed that the proposed likelihood in \citet{distiller2015singlecatch} relies on the assumption that capture histories are independent across animals, which does not hold for single-catch data, and we presented an exact and computationally tractable likelihood. 
The resulting maximum likelihood estimators performed well in simulations.

Concerning removal surveys, \citet{efford2009density} wrote that their data ``are not useful for fitting movement based models such as we describe, because in the absence of recaptures we have no information on the scale of movements.'' In terms of our notation, they are arguing that the cumulative capture hazard $\Lambda(t; d)$ at time $t$ and activity center distance $d$ is not identified and hence that the abundance $N$ and density $f$ are not identified. 
Later work or discussion has not revisited spatially explicit removal estimation, at least as far as the authors are aware. 
We have shown, however, that such data can support spatially explicit inference without recaptures: our general likelihood includes removal surveys as a special case, and \Cref{prop:removal-identification} establishes identification of abundance, while the subsequent discussion suggests that selective depletion may also provide information about spatial detection scale. In simulations, the proposed spatial estimator essentially removed the abundance bias of the non-spatial Zippin estimator, and its profile-likelihood intervals generally attained close to nominal coverage, although poor Wald coverage suggested weak identification. %

In addition to addressing these two open problems in statistical ecology, 
the proposed framework also gives the first maximum likelihood estimator for multi-catch detectors with capture times. This flexible framework can accommodate survey methods and detector types beyond those considered here. 
In future work, we hope to consider nonparametric analysis and further study identification in the spatial removal method. 
We also plan to study data in which capture times are observed only to lie within an occasion so are interval-censored. Although we can readily see that the resulting likelihood is an integral of the likelihood in \Cref{thm:lik} over the possible capture times, there are computational difficulties in survey methods with trap competition such as single-catch and removal. 
In practice, though, we encourage surveyors to consider recording capture times since the extra information is useful for statistical modeling \citep{tsiatis2006semiparametric,haines2023exact}.

\section*{Acknowledgement}

The authors thank Chris Sutherland for helpful conversations. 
OpenAI's ChatGPT was used to assist with numerical implementation, including drafting and debugging code, and proofreading the manuscript.

\bibliographystyle{apalike}
\bibliography{refs.bib}

\end{document}